\documentclass[aps,amsfonts,amssymb,twocolumn,amsmath,superscriptaddress,prb]{revtex4-2}[12pt]

\usepackage{graphicx}%
\usepackage{dcolumn}%
\usepackage{bm}%

\usepackage{xcolor}
\usepackage[pdftex,colorlinks=true,linkcolor=blue,citecolor=blue,urlcolor=blue,filecolor=blue,breaklinks=true]{hyperref} %
\usepackage{url}
\usepackage{braket}

\usepackage{setspace}

\usepackage{textcomp}
\usepackage{gensymb}
\usepackage{mathtools}
\usepackage{cleveref}
\usepackage{siunitx}
\usepackage[final]{changes}

\DeclareSIUnit{\angstrom}{\text {Å}}
\DeclareSIUnit{\gauss}{\text {G}}
\usepackage{float}
\usepackage{comment}
\usepackage{verbatim}
\usepackage{gensymb}
\usepackage{array}
\usepackage[version=4]{mhchem}
\usepackage{manfnt}

\begin{document}
\newcommand{\CdAs}{Cd$_3$As$_2$}

\title{Anomalous electronic energy relaxation and soft phonons in the Dirac semimetal \CdAs}
\author{Rishi Bhandia}
\thanks{These two authors contributed equally}

\affiliation{The Institute of Quantum Matter, Department of Physics and Astronomy, The Johns Hopkins University, Baltimore, Maryland 21218, USA}

\author{David Barbalas}
\thanks{These two authors contributed equally}

\affiliation{The Institute of Quantum Matter, Department of Physics and Astronomy, The Johns Hopkins University, Baltimore, Maryland 21218, USA}

\author{Run Xiao}
\affiliation{Department of Physics, The Pennsylvania State University, University Park, Pennsylvania 16802, USA}
\affiliation{The Institute of Quantum Matter, Department of Physics and Astronomy, The Johns Hopkins University, Baltimore, Maryland 21218, USA}

\author{Juan R. Chamorro}

\affiliation{Department of Chemistry, Johns Hopkins University, Baltimore, Maryland 21218, USA}
\affiliation{The Institute of Quantum Matter, Department of Physics and Astronomy, The Johns Hopkins University, Baltimore, Maryland 21218, USA}

\author{Tanya Berry}

\affiliation{Department of Chemistry, Johns Hopkins University, Baltimore, Maryland 21218, USA}
\affiliation{The Institute of Quantum Matter, Department of Physics and Astronomy, The Johns Hopkins University, Baltimore, Maryland 21218, USA}

\author{Tyrel M. McQueen}
\affiliation{The Institute of Quantum Matter, Department of Physics and Astronomy, The Johns Hopkins University, Baltimore, Maryland 21218, USA}
\affiliation{Department of Chemistry, Johns Hopkins University, Baltimore, Maryland 21218, USA}
\affiliation{Department of Materials Science and Engineering, The Johns Hopkins University, Baltimore, Maryland 21218, United States}

\author{Nitin Samarth}
\affiliation{The Institute of Quantum Matter, Department of Physics and Astronomy, The Johns Hopkins University, Baltimore, Maryland 21218, USA}
\affiliation{Department of Physics, The Pennsylvania State University, University Park, Pennsylvania 16802, USA}
\affiliation{Department of Materials Science and Engineering, The Pennsylvania State University, University Park, Pennsylvania 16802, USA}

\author{N. P. Armitage}
\thanks{npa@jhu.edu}
\affiliation{The Institute of Quantum Matter, Department of Physics and Astronomy, The Johns Hopkins University, Baltimore, Maryland 21218, USA}

\begin{abstract}
	We have used a combination of linear response time-domain THz spectroscopy (TDTS) and non-linear THz pump-probe spectroscopy to separately probe the electronic momentum and energy relaxation rates respectively of the Dirac semimetal \CdAs.  We find, consistent with prior measurements, that \CdAs\ has an enormous nonlinearities in the THz frequency range. We extract the momentum relaxation rate of \CdAs ~using Drude fits to the optical conductivity. We also conduct THz range 2D coherent spectroscopy and find that the dominant response is a pump-probe signal, which allow us to extract the energy relaxation rate. We find that the rate of energy relaxation decreases down to the lowest measured temperatures.  We connect this to \CdAs's anomalous lattice dynamics, evidence for which is found in its low thermal conductivity and soft phonons in Raman scattering. We believe the lack of a peak in the energy relaxation rate as a function of T is related to the linear in T dependence of the current relaxation at low T; e.g. the phonon scattering is elastic from the lowest measured temperature, 5 K, to at least as high as 120 K.

\end{abstract}

\date{\today}
\maketitle

\section{Introduction}

In the past twenty years, a significant theme in condensed matter physics has been the identification, characterization, and study of a range of materials that possess low energy excitations well-described by massless Dirac fermions. As a consequence of this unusual low energy excitation spectrum, these systems exhibit unique properties in their electronic transport or optical response ~\cite{wehlingDiracMaterials2014, Armitage2018, lvExperimentalPerspectiveThreedimensional2021, andoBerryPhaseAbsence1998}. Although many of these systems are inherently two-dimensional, Weyl or Dirac semimetals are a class of three dimensional materials that host linear band touchings in momentum space ~\cite{youngDiracSemimetalThree2012}.  
These features allow for the exploration of novel materials physics arising from these conditions. Unlike graphene, the three-dimensional nature of topological semimetals allow for the enhancement of light-matter interactions due to their finite thickness, leading to many predictions for their applications in high-speed electronic, optoelectronic, and spintronic devices ~\cite{wangUltrafastBroadbandPhotodetectors2017, chorsiWidelyTunableOptical2020, rizzaClosingTHzGap2022, daPredictionNegativeRefraction2022,arakiLongrangeSpinTransport2021}.

\CdAs~is an archetypal example of a Dirac semimetal ~\cite{borisenkoExperimentalRealizationThreeDimensional2014, liuStableThreedimensionalTopological2014, aliCrystalElectronicStructures2014}. In contrast to other Dirac semimetal candidates, such as $\mathrm{Na_3 Bi}$, \CdAs~is stable under ambient conditions making it ideal for applications in devices. It has an almost cubic symmetry with a lattice superstructure associated with Cd vacancies, which greatly enlarges the unit cell to 80 atoms, but preserves inversion and time-reversal symmetry~\cite{kimPointGroupSymmetry2019}. These symmetries allow for four-fold degenerate Dirac points.  Calculations and photoemission experiments anticipate and find two Dirac points located near the $\Gamma$ point, along the $\Gamma - Z$ direction~\cite{borisenkoExperimentalRealizationThreeDimensional2014}.
 
Before its recognition as a Dirac semimetal, \CdAs\ had already drawn attention for its exceptional mobility, first observed in 1959~\cite{rosenbergCdNoncubicSemiconductor1959}. Recent transport measurements have revealed phenomena like giant magneto-resistance, quantum oscillations, and the quantum Hall effect (QHE) in  \CdAs, alongside observations of the chiral anomaly~\cite{liangUltrahighMobilityGiant2015, akrapMagnetoOpticalSignatureMassless2016, wangAnomalousPhaseShift2016, uchidaQuantumHallStates2017, schumannObservationQuantumHall2018, goyalThicknessDependenceQuantum2018, gallettiAbsenceSignaturesWeyl2019, chengProbingChargePumping2021, xiaoIntegerQuantumHall2022, liNegativeMagnetoresistanceDirac2016}.

The lattice properties of \CdAs\ are also anomalous.  It has one of the lowest lattice thermal conductivities ever measured at room temperature, \qtyrange{0.3}{0.7}{\watt \per \m \per \K}~\cite{chenManipulationPhononTransport2018,yueSoftPhononsUltralow2019,spitzerAnomalousThermalConductivity1966,wangMagneticfieldEnhancedHighthermoelectric2018}.  Crystalline materials with similar atomic masses and structures typically have a lattice thermal conductivity 1-3 orders of magnitude larger~\cite{chenManipulationPhononTransport2018}.  This low thermal conductivity is believed to arise as a consequence of the large unit cell that gives rise to a large number of optical phonons that depress the acoustic phonon velocity and mean free path.  Moreover, an anomalous low temperature softening of a group of zone center Raman active phonons has been found~\cite{yueSoftPhononsUltralow2019}.  It has been proposed that these soft modes and low frequency optical phonons increase the phase space for scattering of the heat-carrying acoustic phonons and are -- in part -- the origin of the low lattice thermal conductivity of Cd$_3$As$_2$~\cite{yueSoftPhononsUltralow2019}.  Despite these observations, little progress has been made on understanding the interplay between \CdAs's unusual electronic and lattice dynamics.  Among other aspects it is essential to understand these effects as they can influence device characteristics under non-equilibrium ``hot-electron'' conditions in various ways.

One method of studying the coupling between electronic and lattice degrees of freedom is non-linear (Terahertz) THz pump-probe experiments~\cite{barbalasEnergyRelaxationDynamics2023, hoffmannTHzpumpTHzprobeSpectroscopy2009, heblingObservationNonequilibriumCarrier2010}. Pump-probe experiments can measure the rate that energy leaves the electronic system via coupling to some external bath~\cite{groeneveldFemtosecondSpectroscopyElectronelectron1995, kabanovElectronRelaxationMetals2008, kabanovElectronelectronElectronphononRelaxation2020}.  This is in contrast to the momentum (or more strictly speaking the current) relaxation rate that is usually measured in conventional transport and optical experiments.  However, most ultrafast pump-probe experiments have utilized near-infrared (NIR) or optical frequency (\qtyrange{1}{3}{\eV}) pump sources, which drive the electronic and lattice sub-systems into regimes very far from equilibrium~\cite{ weberSimilarUltrafastDynamics2017, luUltrafastRelaxationDynamics2017, zhuBroadbandHotcarrierDynamics2017, luTerahertzProbePhotoexcited2018, zhangUltrafastPhotocarrierDynamics2019, reinhofferRelaxationDynamicsOptically2020, zhaiMidinfraredTransientReflectance2020, baoPopulationInversionDirac2022}.  With such high energy excitations, the usual quasi-equilibrium approximations may be invalid, as the energy scales initially excited are many orders of magnitude higher than the energy scales relevant to the low-temperature phenomena of solids.  Hence interpretation usually rests on the assumption of very fast relaxation to a quasi-thermal distribution that is subsequently probed.  In contrast, THz range non-linear spectroscopy is an attractive tool to probe couplings in the low-energy scale (meV) regime.

In this work, we perform both conventional linear time-domain THz spectroscopy and non-linear THz pump-probe on films of \CdAs~to separately measure the momentum and energy relaxation rates. Via conventional linear response time-domain spectroscopy, we observe that the momentum relaxation rate is weakly dependent on temperature up to \qty{300}{K}. Utilizing nonlinear THz measurements, we build on previous studies observing large THz third-harmonic generation in \CdAs, as we measure the temperature-dependent, third-order nonlinear susceptibility $\chi^(3)(\omega)$ over the THz frequency range~\cite{chengEfficientTerahertzHarmonic2020,kovalevNonperturbativeTerahertzHighharmonic2020}. This allows us to more clearly understand dominant nonlinear processes of \CdAs~while elucidating the nature of electron-phonon dynamics via careful examination of pump-probe processes and energy loss from the electron system.   We find that the rate of energy loss from the electronic system decreases as temperature increases from our lowest to highest measured temperatures.  This is at odds with expectation where a peak in the energy relaxation rate is expected at temperature set by the Bloch-Gr{\"u}neisen temperature~\cite{allenTheoryThermalRelaxation1987}.  We connect this surprising behavior to the anomalous phonon dynamics of \CdAs, as evidenced by its low thermal conductivity and the observation of soft phonons in Raman scattering~\cite{yueSoftPhononsUltralow2019,zhangUnexpectedLowThermal2015,wangMagneticfieldEnhancedHighthermoelectric2018,bartkowskiSpecificHeatSingleCrystalline1989}.  We believe the lack of a peak in the energy relaxation rate as a function of T can be connected to the linear in T dependence of the current relaxation e.g. the phonon scattering is dominantly elastic down to the lowest measured temperatures. Both thermal conductivity and electrical resistivity are then dominated by the same scattering mechanisms and one expects that the Wiedemann–Franz law would then be satisfied for \CdAs~over this large range of temperatures~\cite{lavasaniWiedemannFranzLawFermi2019}.

\section{Experimental Details}

The \qty{100}{\nm} \CdAs~(112) thin films were grown via molecular beam epitaxy (MBE) on (111) GaAs substrates with a \qty{100}{\nm} GaSb buffer layer. F{These f}ilms were monitored during the growth process with reflection high-energy electron diffraction (RHEED) to assess the surface quality. See earlier work for more detail on the growth process~\cite{xiaoIntegerQuantumHall2022}. Prior angle-resolved photo-emission spectroscopy (ARPES) on thin films grown under these  conditions have confirmed the presence of Dirac cones while x-ray diffraction has confirmed presence of peaks corresponding to the \CdAs (112)~planes~\cite{yanezSpinChargeInterconversion2021}.

We performed conventional linear response THz measurements on these samples utilizing our custom time-domain THz spectrometer coupled to a helium vapor flow cryostat. The THz pulses were generated using a fiber oscillator laser (Toptica Femtofiber smart 780) that produced a \qty{780}{\nm}, \qty{120}{\fs} pulses at a repetition rate of 80 MHz. These pulses are split into two beams.  One was used to generate a THz pulse using a voltage-biased photo-conductive antenna (PCA), while the other beam was used to gate an unbiased PCA to detect the THz pulse.  By measuring the transmitted THz pulse through the sample and an appropriate substrate reference we can calculate the complex transmission coefficient. Then by inverting the equation $ T\left(\omega\right) = \frac{1 + n_{sub}}{1 + n_{sub} +Z_0 d \sigma\left(\omega\right)} \mathrm{exp}\left[ i\frac{\omega }{c} \left( n_{sub} - 1 \right) \Delta L\right]$ we can obtain the complex conductivity $\sigma\left(\omega\right)$ of the film. Here $T\left(\omega\right) $ is the measured complex transmittance of the thin film, $n_{sub}$ the index of refraction of the substrate, $d$ the thin film thickness, and $\Delta L$ the factor to account for differences in thicknesses between substrate and sample.

\begin{figure*}
\begin{center} 
	\includegraphics[width = 0.8\textwidth]{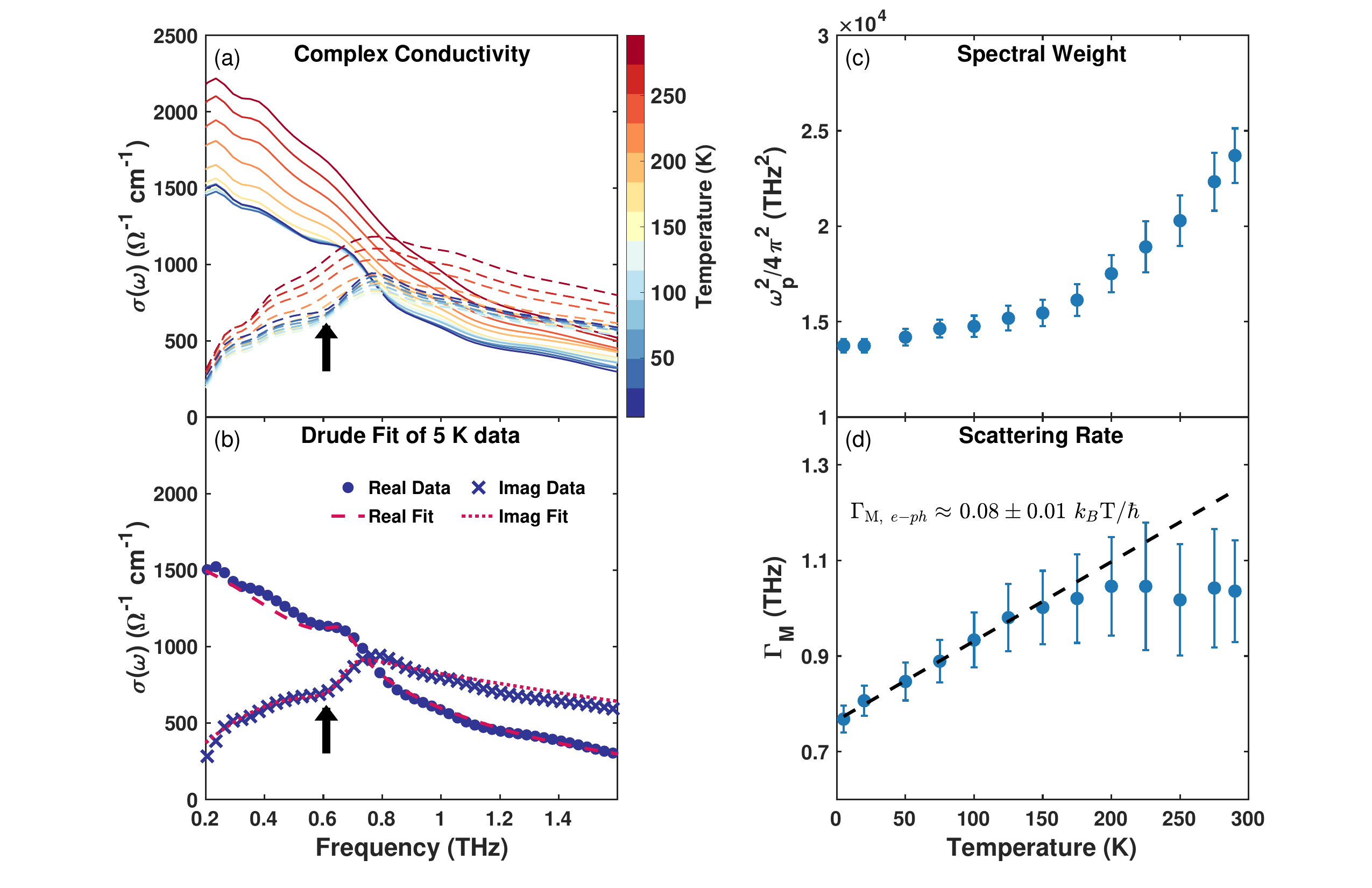}
\end{center}
	 \caption{ (a) THz complex conductivity of a \qty{100}{\nm} thin film of \CdAs. The real (imaginary) optical conductivity is plotted as solid (dashed) lines for temperatures from \qtyrange{5}{290}{\K}. The arrow indicates a feature in the conductivity corresponding to an optical phonon. (b) An example fit (markers) to the model described in the text (dashed lines) (c) The fitted spectral weight of the electronic Drude term. (d) The scattering rate, $\Gamma_M$, obtained from the Drude fits. Dotted line is a linear fit emphasizing the linear in T dependence from \qty{5}{\K} to \qty{120}{K}, corresponding to elastic scattering from phonons. From this fit we can estimate the electron-phonon contribution to the scattering rate as $0.08 \pm 0.01\ k_B \mathrm{T}/h$.}\label{fig:Conductivity}
\end{figure*}

For the nonlinear THz spectroscopy, we conducted experiments in a now standard THz 2D coherent spectroscopy geometry with two \ce{LiNbO3} THz sources as described elsewhere~\cite{mahmoodObservationMarginalFermi2021, barbalasEnergyRelaxationDynamics2023, woernerUltrafastTwodimensionalTerahertz2013}.  At its maximum, the incident field strength was \qty{40}{\kilo \V \per \cm} per pulse with a center frequency of 0.6 THz. To measure the nonlinear response, we used a differential chopping scheme that allow us to measure the transmitted pulses, $E_{A}$ and $E_B$ separately, and then pulses $E_{AB}$ transmitted together. From this, we can calculate the nonlinear response via subtraction, $E_{NL} = E_{AB} - E_A - E_B$. We sample the response in time, $t$, and control the delay between the pulses, $\tau$, with a pair of delay stages. The incident electric field strength is controlled with a pair of wire-grid polarizers. If the THz pulse energy is absorbed entirely into the electronic system and can be considered to go entirely into a thermal distribution the heating at room temperature (\qty{300}{K}) of the electronic subsystem is negligible, but can become significant at low temperatures. We elaborate on the impact of this heating on the interpretation of our pump-probe experiments further below.

\section{Analysis and Results}

The complex optical conductivity measured at different temperatures for a representative thin film is shown in Fig.~\ref{fig:Conductivity}(a).  One sees a large Drude-like peak that arises from free-charge carrier motion. At lower temperatures, a very small 0.6 THz optical phonon feature that was resolved in previous studies becomes apparent~\cite{chengLargeEffectivePhonon2020}. The temperature dependent conductivity is roughly what we expect for a high mobility semimetal. A larger width indicates increased carrier scattering at high temperature.  The increased area and overall scale of the conductivity curves indicates slightly larger plasma frequency at higher temperatures. We can fit the optical conductivity to the standard expression for the Drude model with another term for the optical phonon

 \begin{align}
 	 \sigma(\omega) = \epsilon_0 \left[\frac{-\omega_p^2}{i\omega - 2 \pi \Gamma_M} - i(\epsilon_\infty - 1)\omega\right] + \sigma_{\text{ph}}(\omega),
 \end{align} 
 where $\omega_p$ is the Drude plasma frequency, $\Gamma_M$ is the current relaxation rate, and $\epsilon_\infty$ is the high frequency dielectric constant, which accounts for the effects of higher band inter-band transitions on the low frequency dielectric constant. $\sigma_{\text{ph}}$ is a Fano-shaped Lorentz oscillator that accounts for the screened optical phonon that appears at lower temperatures near 0.6 THz~\cite{damascelliOpticalSpectroscopyQuantum1999, homesVibrationalAnomaliesFe2018}. Panel (b) in Fig. \ref{fig:Conductivity} shows an example fit. This phonon had been seen in previous studies~\cite{chengLargeEffectivePhonon2020}, but appears much weaker here due to the larger carrier density of these films and attendant screening effects.  The results of these fits are shown in Figs. \ref{fig:Conductivity} (c) \& (d). The scattering rate, $\Gamma_M$, is plotted as a function of temperature in Fig. \ref{fig:Conductivity}(d). Formally, this rate is the current or transport relaxation rate ~\cite{lavasaniWiedemannFranzLawFermi2019}. For materials with sufficiently isotropic dispersion, it is equivalent to the momentum relaxation rate.  As one would expect, this rate increases with increasing temperature in all the \CdAs~films measured, however its explicit dependence can be noted as anomalous.  One expects the electron-phonon interaction to give a low-temperature scattering rate that goes as T$^3$ - T$^5$ depending on the degree to which large angle scattering is important. We instead observe that $\Gamma_m \propto T$ up to \qty{\sim 120}{\K}, as shown in Fig. 1(d), above which the T dependence gets weaker.

\begin{figure*}
\begin{center}
	\includegraphics[width = 0.9\textwidth]{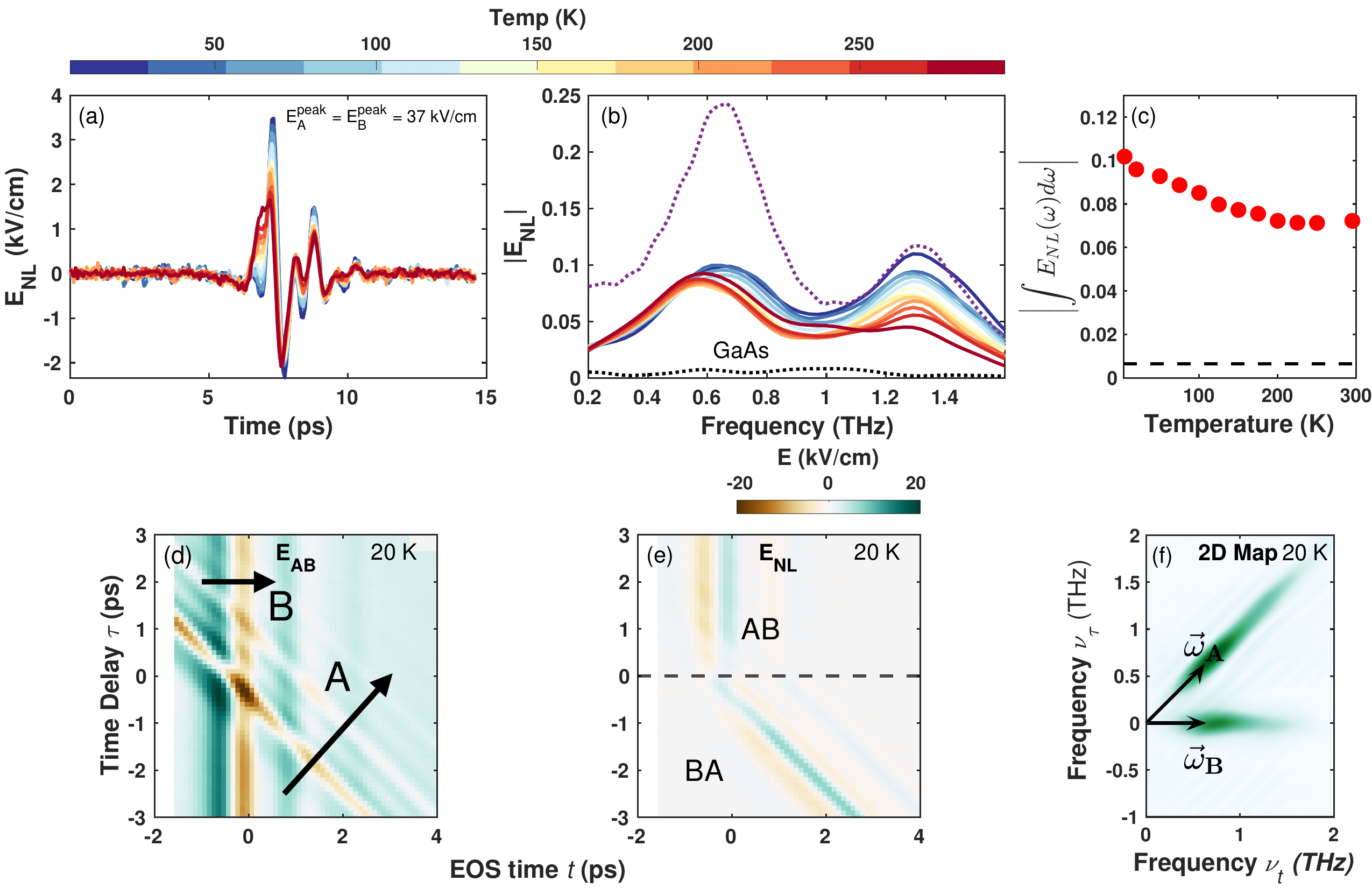}
\end{center}
	 \caption{The nonlinear THz spectra of the 100 nm \CdAs\ thin film. (a) Time trace of the nonlinear signal $E_{NL}(t,\tau = 0)$ (b) The Fourier transform of the data in panel (a). The black dotted line is the $\tau = 0$ nonlinear response of the bare \ce{GaAs} substrate. The purple dotted line is the input pulse spectrum divided by a factor of 7.5 to be made visible on the plot.  (c) The nonlinear spectra from panel (b) integrated over the range \qtyrange{0.3}{1.4}{\THz}. The dotted black line gives the intensity of \ce{GaAs} nonlinear spectrum integrated over the same frequency range. (d) Time traces of $E_{AB}(t,\tau)$ (e) The nonlinear component of panel (d), $E_{NL}(t,\tau)$. (f) A 2D THz spectra of $\left| E(\nu_{\tau},\nu_t))\right|$ obtained by taking at 2D FFT of $E_{NL}(t,\tau) $. The only nonlinear process we observe is pump-probe.}\label{fig:nonLinearTimeTrace}
\end{figure*}
 
As the plasma frequency ($\omega_p$) varies inversely with the effective mass, we can attribute the plasma frequency's measured temperature dependence to the temperature dependence of the effective mass. Since $\omega_p \propto E_F$ for electrons with linear dispersion in three dimensions, when comparing to prior measurements of the plasma frequency in magento-optic experiments~\cite{chengLargeEffectivePhonon2020}, we can estimate the Fermi energy to be \qty{\sim 0.2}{\eV} above the Dirac point. This is consistent with the Fermi energy measured via ARPES in \CdAs~thin films grown under similar conditions~\cite{yanezSpinChargeInterconversion2021}. It is interesting to note the approximately T$^2$ dependence of the plasma frequency (with an offset) is similar to what is predicted for linearly dispersing systems in general~\cite{ashby2014chiral}.  However, note that \CdAs~does have large parabolic contributions to its dispersion only 10s of meV away from its Dirac points, so any close comparisons would have to take into account details of the band structure~\cite{akrapMagnetoOpticalSignatureMassless2016}.

To determine the overall scale of the nonlinear response, we first conducted experiments where we set the time delay between the two pulses to $\tau=0$ and then measured $E_{NL}\left(t,\tau = 0\right)$ as discussed above. In Fig. \ref{fig:nonLinearTimeTrace}(a), we plot the time trace of the nonlinear response at different temperatures. Here the plotted electric field is the electric field strength at the sample position. As shown, the nonlinear response increases monotonically as we cool to lower temperatures.  To confirm that the nonlinear response is intrinsic to the \CdAs, we also measured the nonlinear response of $\mathrm{GaAs}$ as a function of temperature. In Fig. \ref{fig:nonLinearTimeTrace}(b) the Fourier transform (FFT) of the nonlinear response is plotted at different temperatures.  We observe that the nonlinear response of the sample is far larger than the nonlinear response of the substrate (thin dashed black line) confirming that the source of these optical nonlinearities is the \CdAs~thin film. The purple dotted line is the input pulse, $E_{AB}$, divided by a factor of 5 to be made visible on the plot.

\begin{figure}[t]
\begin{center}
	\includegraphics[width = 0.47\textwidth]{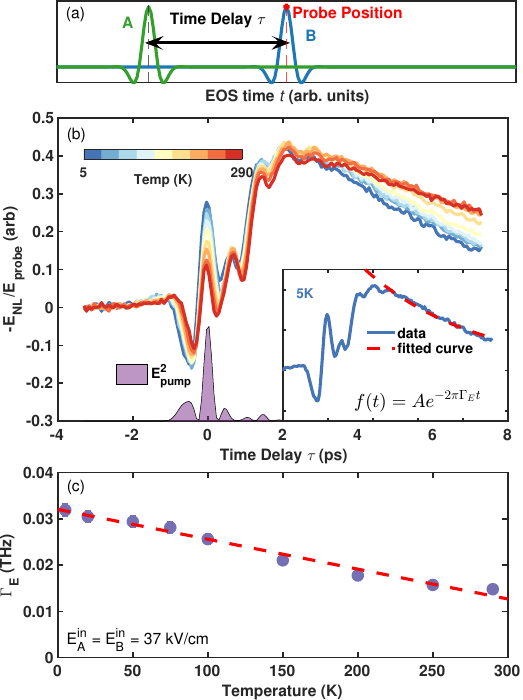}
\end{center}
	 \caption{Nonlinear THz pump-THz-probe time traces of a representative 100 nm \CdAs\ thin film. (a) A schematic of our experiment. (a) Pump-probe time traces as a function of temperature. Note that the decay rate changes visibly, with the slope becoming steeper at lower temperatures. The purple shaded area is the time profile of $E^2_{pump}(\tau)$ inside the sample in arbitrary units. $\int E^2_{pump}(\tau) d\tau \propto Q_{film}$, the heat energy deposited in the electronic sub-system.  The inset is an example fit of a decay to extract the energy relaxation time. (c) Measured energy relaxation rates $\Gamma_E$ as a function of temperature. }\label{fig:PumpProbe}
\end{figure}

To identify the source of the nonlinear signal, we can construct a 2D coherent spectrum by sweeping the time delay between the two intense THz pulses and the electro-optic sampling time. The results are shown in Fig. \ref{fig:nonLinearTimeTrace}(d). We plot the nonlinear component of the THz response in Fig. \ref{fig:nonLinearTimeTrace}(e) and take a 2D FFT of it to obtain a 2D spectra as shown in Fig. \ref{fig:nonLinearTimeTrace}(f). Here we have two different ``pump-probe" contributions appearing as streaked peaks.  The two separate peaks in the 2D plot correspond to the two different possible time orderings of the A and B pulses; the diagonal (horizontal) is where pulse B (pulse A) acts as the pump pulse and pulse A (pulse B) acts as the probe, known as the BA regime (AB regime).  The position they appear in the 2D spectra can be understood according to frequency vectors $\vec{\omega}_{AB:PP} = \vec{\omega}_{A} - \vec{\omega}_{A} + \vec{\omega}_{B} = \vec{\omega}_{B} $ or $\vec{\omega}_{BA:PP} = \vec{\omega}_{B} - \vec{\omega}_{B} + \vec{\omega}_{A} = \vec{\omega}_{A}$ and hence follow from processes that go as $E_A^* E_A E_B$  and $E_B^* E_B E_A$~\cite{mahmoodObservationMarginalFermi2021}.  These are pump-probe contributions to the nonlinear response.  As $E_A^* E_A$ and $E_B^* E_B$ are intensities, the dependence of the response as a function of the interpulse separation $\tau$ is a measure of how fast the system recovers from pump.  Such nonlinear pump-probe processes are the expected dominant nonlinear process in metals~\cite{confortiDerivationThirdorderNonlinear2012}. 

In general, the pump-probe processes driving these nonlinearities can be understood as the THz pump (here $E_A$), creating a Fermi surface perturbation, which is then probed by $E_B$ as it relaxes back to equilibrium. To get a very rough feeling for the size of these effects, we can consider heating due to the high-field THz pulses incident on the sample. As nonlinearities are, in general, small as compared to the linear effects, we make an estimate of the energy absorbed per unit volume by $Q = \int_0^T \sigma_1(\qty{0.6}{\THz})  [E^{inside}_{A} (t) ]^2 dt$, where $T$ is the pulse length ($\sim$1 ps). We use $\sigma_1( 0.6 \; \mathrm{THz})$ as an estimate of the relevant conductivity as it is the frequency of maximum THz field.   $E^{inside}_{A}$ can be estimated from $E_{A}$ as discussed in the SM of Ref.~\cite{katsumiRevealingNovelAspects2023} to be at a maximum drive field of  \qtyrange{\sim7}{9} {\; \kV \per \cm}.  As our measured THz conductivity spectra is dominated by the intra-band Drude term, the deposited energy goes into raising the temperature of a quasithermal electronic distribution.  One can estimate the rise in effective temperature of the electrons (using the measured electronic contribution to the heat capacity of $C_e = \gamma T$ with $\gamma = \qty{5.33}{mJ.mol^{-1}.K^{-1}}$ ~\cite{zhuBroadbandHotcarrierDynamics2017, wuLargeAnisotropicLinear2015}) as \qty{58}{K} at \qty{10}{K} and \qty{5.8}{K} at \qty{100}{K}.   Hence the perturbation is large at the lowest temperatures for the largest drive fields, but as we will note we see the same qualitative temperature dependencies for the lower drive fields that deposit only 1/10 as much energy.

\begin{figure}[h]
\begin{center}
	\includegraphics[width = 0.47\textwidth]{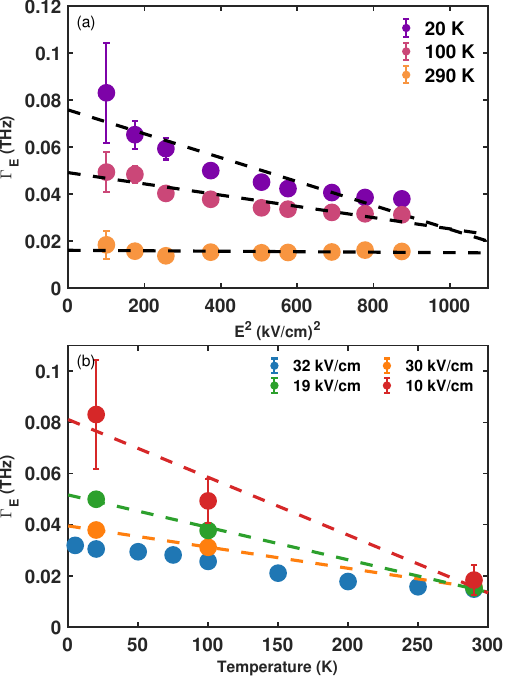}
\end{center}
	 \caption{Electronic energy relaxation rates as a function of fluence and temperature. (a) Energy relaxation rates for a single 100 nm thin film at different applied electric field strengths at three different fixed temperatures. The dashed lines are a guide to the eye. (b) Energy relaxation rates as a function of temperature at several fixed field strengths. The dashed lines are again a guide to the eye. }\label{fig:EnergyRelaxationRates}
\end{figure}

A schematic of our pump-probe experiments is shown in Fig. \ref{fig:PumpProbe}(a). The probe time $t$, is fixed as the time delay between the pulses, $\tau$, is varied. Pump-probe time traces taken at different temperatures are shown in Fig. \ref{fig:PumpProbe}(b).  We normalize the plotted nonlinear signal, $E_{NL}$ by the probe field ($E_B$) to account for the temperature dependent conductivity of the film. A factor of $-1$ is included to account for the fact the pump decreases the transient transmittance. For times after both pumps have arrived, one sees two distinctly different regions of behavior.  At early times, there is interference between the A and B pulses.  As one can see from Fig. \ref{fig:nonLinearTimeTrace}(e), this comes from a region of order the pulse duration ($\sim 2$ ps) around $\tau  \sim - t $ where there is ambiguity about the time ordering of the pulses.  This leads to oscillations in this time region that are due to the pump-probe processes, but from when B pulse comes before A. For times $\tau \gg - t $, the signal shows decaying behavior.  As discussed in Ref.~\cite{barbalasEnergyRelaxationDynamics2023}, after an E$^2$ perturbation the decay at long times is expected to be largely governed by the rate of energy relaxation e.g. the rate that energy leaves the electronic systems.  By fitting to a model of exponential decay ($\propto e^{-2 \pi \Gamma_E  t} $) at times well after the pulse (as shown in Fig. \ref{fig:PumpProbe}(b)), we obtain the energy relaxation rate as shown in Fig. \ref{fig:PumpProbe}(c) at different temperatures.  We find that $\Gamma_E$ increases as temperature decreases over the whole measured temperature range (Fig. 3(c)).  This $\Gamma_E$ is associated with energy leaving the electronic subsystem via coupling to non-optically active degrees of freedom~\cite{gloriosoJouleHeatingBad2022}. One generally expects the energy relaxation rate to be less than the momentum relaxation rate, as the latter is essentially a back-scattering weighted average over all scattering processes, whereas $\Gamma_E$ is only sensitive to a small minority of inelastic processes that remove energy from the electronic system. In a high mobility semimetal, we expect it to arise from mainly electron-acoustic phonon scattering.   For low frequency phonons, it is expected that at higher temperatures are they are excited to such a degree that they are as absorbed as often as they are emitted and are hence inefficient sinks of energy.
   
It is interesting to compare the observed rates to expectation.  The first derivation of energy relaxation rates was by Allen~\cite{allenTheoryThermalRelaxation1987} based on a model of a Debye model for the acoustic phonons (with subsequent refinements~\cite{groeneveldFemtosecondSpectroscopyElectronelectron1995, gloriosoJouleHeatingBad2022}). In this  ``two-temperature'' model, one considers the system after optical excitation as having electrons that quickly thermalize to a temperature T$_e$ and then over longer time scales lose energy to the lattice that is at a lower temperature T$_L$.  At temperatures well above the Bloch-Gr{\"u}neisen scale, $\Gamma_E$ is expected to have an increasing dependence with decreasing temperature that goes as

\begin{align}
 \Gamma_E = \frac{ 6 \hbar \lambda \left< \omega^2\right>}{k_B \mathrm{T}_e}, 
\end{align}\label{eq:AllenHighTempRelaxation}
where T$_e$ is the temperature of the electronic subsystem and $ \lambda \left<  \omega^2\right>$ is the second moment of the Eliashberg function given by $\lambda \left<  \omega^2\right> = 2 \int_0^\infty d\Omega \left[\alpha^2 F(\Omega) / \Omega \right] \Omega^2 $~\cite{allenTheoryThermalRelaxation1987, groeneveldFemtosecondSpectroscopyElectronelectron1995, gloriosoJouleHeatingBad2022} that quantifies the coupling to acoustic phonons.  In a conventional metal, the Bloch-Gr{\"u}neisen temperature is T$_{BG} = 2 k_F v_s \hbar/ k_B $ (where $v_s$ is the speed of sound).  As the temperature is lowered below  $\sim 0.3 \mathrm{T}_{BG}$~\cite{gloriosoJouleHeatingBad2022}, $\Gamma_E $ starts to decrease and is expected to crossover to a dependence that goes as T$_e^3$ with a coefficient close to the expectation for the acoustic phonon contribution to the electron-hole two-particle self energy.  The data is at odds with these simple expectations as  $\Gamma_E $ is an increasing function with decreasing T over the whole range.

In Figs.~\ref{fig:EnergyRelaxationRates}(a) - (b), we plot the energy relaxation rates as a function of field strength (normalized to the maximum field \qty{40}{\kV\per\cm}) at different temperatures and as a function of temperature at different fields.  At room temperature the relaxation times, \qty{290}{\K}, show no dependence on the pump electric field. However, at lower temperatures we see a dependence that shows enhanced decay rates at the lowest temperatures and lowest drive fields.   This dependence of the rate on field is indicative of being out of the perturbative $\chi^{(3)}$ regime at the lowest temperatures. From the pump probe time traces, we can estimate that the induced change in conductivity at the highest fluence is of the same scale as the equilibrium conductivity itself, \qty{\sim 1200}{\per\ohm\per\cm} consistent our observation of a non-perturbative response \footnote{These estimates were obtained with the expression $\Delta \sigma(\tau) \approx \frac{n_{sub} + 1}{d Z_0 T}\left(\frac{- E_{NL}(\tau)}{E_{AB}(\tau) - E_{A}}(\tau)\right)$ }.

From \qtyrange{150}{290}{\K}, the measured energy relaxation rate is qualitatively consistent with conventional expectations.  $\Gamma_E$ does shows the expected increasing dependence with decreasing temperature, but note that the dependence is far from $1/T$. Quite anomalously, the dependence is monotonic down to scales well below any reasonable estimate for this material's T$_{BG}$.  In a conventional metal, T$_{BG}$ is approximately the Debye temperature as 2$k_F$ is of order a reciprocal lattice vector.  The Debye temperature of \CdAs~is estimated to be approximately \qty{100}{\K} via the specific heat~\cite{decombarieuSpecificHeatCadmium1982} or 187 K via sound velocity~\cite{wangUnusualThermodynamicsLowenergy2022}. These estimates of the Debye temperature can serve as a first estimate for the temperature range over which $\Gamma_E$ is expected to increase with decreasing temperature. Of course, \CdAs~is a semimetal and can -- in principle -- have a $k_F$ that is much smaller than a reciprocal lattice vector.  Taking our earlier estimate of the Fermi energy to be \qty{\sim0.2}{\eV}, this corresponds to an estimated $k_F$ of \qty{\sim0.08}{\per\angstrom} under the modified Kane model for \CdAs~\cite{chengProbingChargePumping2021,wangThreedimensionalDiracSemimetal2013}. This is \qty{16}{\percent} of the reciprocal lattice scale and so suggests the Fermi surface is not small enough to account for the discrepancy by itself.  A far more likely source of the anomalies is that the phonon spectrum in Cd$_3$As$_2$ is found to be extremely non-Debye like.  As mentioned, Cd$_3$As$_2$'s unit cell is extremely large, incorporating 80 atoms in a unit cell. This leads to an unusually small Brillouin zone.  Materials with small Brillouin zones tend to have anomalous phonons with many optical modes and low velocity acoustic ones~\cite{chenManipulationPhononTransport2018}.  In this regard an examination of the specific heat~\cite{decombarieuSpecificHeatCadmium1982} suggests the phonon spectrum is strongly non-Debye-like, and has a number of anomalous and very soft low frequency phonons that have also been observed in Raman~\cite{yueSoftPhononsUltralow2019}.  Evidence for softening of phonon modes below \qty{100}{\K} due to the proximity of the Dirac semimetal state to a Kohn anomaly has been found in Raman scattering~\cite{yueSoftPhononsUltralow2019}. This softening leads to an increased phase space for phonon-phonon scattering and low phonon velocities and high damping.   Irrespective of the origin, this reinforces the notion that the Debye temperature is not the relevant temperature scale to characterize the phonon spectrum in \CdAs.   Using a sound velocity of \qty{\sim 2000}{\meter \per \second} ~\cite{liangUltrafastOpticalProbe2022,  wangUnusualThermodynamicsLowenergy2022} and the $k_F$ given above, one can estimate T$_{BG}/3$ to be \qty{\sim 40}{\K}, which is not so far from the lowest temperature we measure down to in Fig.~\ref{fig:PumpProbe}(c). Nevertheless the phonon spectrum is likely to be so non-Debye as to require explicit detailed calculations to find any agreement with experiment.

In the conventional theory of energy relaxation coupling of low energy acoustic phonons to electronic degrees of freedom are considered~\cite{allenTheoryThermalRelaxation1987}.  Of course optical phonons can also serve as a sinks for energy.   In most materials, their energies are too high to be relevant to relaxation of a quasi-thermal distribution.   However in addition to anomalous acoustic phonons Cd$_3$As$_2$ also has anomalous very low energy optical phonons that could also carry off energy.   Optical phonons are generally expected to give an exponentially activated form for $\Gamma_E$ at low temperature and the same $1/T$ dependence at high temperature~\cite{gloriosoJouleHeatingBad2022}.  So it is not even clear that their consideration would give greater understanding of the present data.   Nevertheless it would be interesting to do a detailed calculation using an experimentally determined phonon spectrum and calculate the energy relaxation rates taking into account both acoustic and optical phonons.

In the conventional theory, the peak in $\Gamma_E $ is found at a temperature above which the electron-phonon scattering is largely elastic.  As no peak is found, the electron-phonon scattering can be inferred to be elastic for all measured temperatures and thus the Wiedemann-Franz law should be obeyed  for the measured temperature ranges.   Tempeature dependent elastic scattering is largely consistent with the fact that the current scattering rate is linear in T from the lowest measured T all the way up to \qty{120}{\K}.   We note that with the previous measurements of the dimensionless electron-phonon coupling constant of $\lambda \approx $ 0.09 (for optical phonons), the scattering rate in the elastic regime is predicted to be $2 \pi \lambda k_B \mathrm{T}/h \approx 0.56 k_B\ \mathrm{T}/h $.   Although this is much larger than the approximately $0.08 \pm 0.01\ k_B\mathrm{T}/h $ found above, it is notable that the linear T dependence is found in this regime. The numeric discrepancy may be considered unsurprising due to the extremely non-Debye character of the phonon spectrum and different phonon branches probed.

\section{Conclusion}
In this work we have performed linear and nonlinear THz spectroscopy on the Dirac semimetal \CdAs. 
We have found energy relaxation rates that are in magnitude far smaller then the momentum relaxation rates.  The energy relaxation rate shows a temperature dependence qualitatively similar to that expected for a metal well above T$_{BG}$, but surprisingly exhibits this dependence to the lowest measured temperature, which are far lower than any conventional estimates for the Bloch-Gr{\"u}neisen temperature.  We find an electric field dependence that is consistent with this behavior and determine we are in a regime of temperature that gives an energy relaxation with an extremely low effective T$_{BG}$. This extremely low T$_{BG}$ may stem from the low effective sound velocity induced by large number of optical phonon modes created by \CdAs's large unit cell~\cite{chenManipulationPhononTransport2018}.   We believe the low effective T$_{BG}$ corresponds to a regime of elastic electron-phonon scattering over the entire measured temperature range and also manifests itself in the linear in T dependence of the current dependent scattering rate that is observed up to \qty{120}{\K}.
The nonlinear response we measure in the THz range is strongly non-perturbative, consistent with prior measurements. However, we find that the origin and scale of the nonlinear signal in \CdAs~is directly connected to the heating and then subsequent relaxation of the electronic subsystem. We can observe this via the temperature and fluence dependence of our nonlinear response, which is enhanced as the heating of our electronic subsystem increases at lower temperature. This is substantiated by the lack of fluence dependence in our measured energy relaxation rates at room temperature, while at lower temperatures we observe strong dependence between pump fluence and relaxation rates.

\acknowledgments

Work at JHU and PSU was supported as part of the Institute for Quantum Matter, an Energy Frontier Research Center funded by the Office of Basic Energy Sciences of the Department of Energy, under grant no. DE-SC0019331.  Instrumentation development at JHU was supported by the Gordon and Betty Moore Foundation EPiQS Initiative Grant GBMF-9454.

\bibliography{Cd3As2_corrected_compounds}

\end{document}